\documentstyle[pre,aps,epsfig,eqsecnum]{revtex}

\begin{document}
\draft  

\title{Random deaths in a computational model for age-structured populations}
\author{J.S. S\'a Martins$^{\dag}$ and S. Cebrat$^{\ast}$}
\address{$\dag$ Colorado Center for Chaos and Complexity, CIRES, CB 216,  
University of Colorado, Boulder, Colorado, USA 80309}
\address{$\ast$ Institute of Microbiology, University of Wroc{\l}aw, ul.S. 
Przybyszewskiego 63/77, 51-148 Wroc{\l}aw, Poland}
\date{\today}
\maketitle

\begin{abstract}
The concept of random deaths in a computational model for population dynamics 
is critically examined. We claim that it is just an artifact, albeit useful, 
of computational models to limit the size of the populations through the use 
of the so-called Verhulst factor and has no biological foundation. 
Alternative implementations of random deaths strategies are discussed and 
compared.
\end{abstract}

\section{Introduction}
The concept of random deaths, an inconspicuous element of computational 
models designed for studies of population dynamics, has caused some 
controversy lately. Its main drawback comes from a strictly biological 
perspective: with very few exceptions, such as some human fishing practices, 
it is doubtful that any significant proportion of the deaths in real 
populations come from random causes. The usual justification for the 
introduction of the concept relates it to limitations on the size of the 
population caused by the finite amounts of food and space provided by the 
environment, and random deaths would be the outcome of intra-species 
competition for these resources. A similar argument also holds when one 
considers the action of predators as a limiting factor for the size of the 
population, and once again random deaths are summoned to account for the 
final effect. 

In both cases, the random component of the deaths they cause is hardly 
significant, if at all. Genetic fitness should account for the success or 
failure in dealing with both constraints. The need for the concept of random 
deaths stems from the limited capabilities of present day models to encompass 
all of the relevant features of the life and death cycles of populations. It 
is also usually the only density-dependent regulatory mechanism, known to 
exist in real populations \cite{cworth}, provided by these models. In most 
cases, the concept appears under the form of the logistic, or Verhulst, 
factor; it has an important role in keeping the size - or, perhaps more 
precisely, the growth \cite{raab} - of the populations simulated in computer 
models within bounds through the killing of a fraction of the population due 
to causes not dealt with by that particular model. Its importance is thus 
undeniable, if one intends to study the problems of population dynamics 
through computer modeling. However, one must be sure that its inclusion does 
not affect the particular feature that the model intends to capture; 
otherwise, the results can be misleading. In this paper we will show that 
this is exactly the case for a very popular model in use today, the 
bit-string Penna model \cite{penna} used in the simulation of the dynamics 
of age-structured populations. This is done by comparing the results obtained 
with the use of two different versions of the implementation of the Verhulst 
factor.

Once the effects of the Verhulst factor are recognized, we have to choose 
between the two alternative implementations. The authors would be very glad 
not to have to deal with random deaths in their future work in the field, but 
could see no clear cut way of dispensing it altogether. The situation seems 
to be similar to the one faced by the classical statistical mechanics 
treatment of e.g. gases, in which events are described as random if their 
causes are too complicated to be analyzed precisely. We present some 
evolutionary arguments, together with the results of simulations of the 
coevolution of the two alternate populations, that favor, in the general 
case, one of the implementations. 

\section{The Verhulst factor}

The usual implementation of random deaths in models for population dynamics 
is through the use of the Verhulst factor. This is a time-dependent death 
probability due to causes that are not dealt with by the dynamic rules of 
the model. Its usual form is $V(t) = N(t)/Popmax$, where $N(t)$ is the total 
population at the beginning of time step $t$ and $Popmax$ is a parameter. In 
the original version of the Penna bit-string model, at each time step every 
individual in the population, irrespective of present age or programmed death 
age, can be killed, with a probability $V(t)$. In what follows, we will code 
this strategy VA, for Verhulst equal for All. Because of the random nature 
of this rule, the genome space of the population is homogeneously sampled; 
well-fitted and ill-fitted individuals die with equal probability. As already 
pointed out in the Introduction, we can see no biological justification for 
this randomness. In a competition for limited resources and in the struggle 
to escape predation, better fitted organisms would most certainly be killed 
with a smaller probability. From another point of view, the economics of 
such random deaths would certainly be too costly, since it would expose 
equally to a premature killing individuals for which different investments had 
been made. Since the only valid reason for the Verhulst factor to still be 
part of the overall dynamic rules lies in the need to keep the size of the 
population limited, it would be in principle desirable to find an alternative 
and less costly strategy population-wise for its implementation.

Such an alternative has recently been suggested \cite{cebrat}. Instead of 
acting as a random death probability for all the population, the Verhulst 
factor acts only on the individuals whose genomes have not been tested by 
the environment yet, the newborn (VB). In fact, a similar strategy for the 
Verhulst factor was adopted by a recent simulation of the Penna model on a 
lattice \cite{makowiec}. From the economics of the population, this is 
clearly a better choice, since little investment is wasted. From the 
biological perspective, although it is not yet the most faithful 
representation of the real natural processes, it has an advantage since the 
genome of the newborn is on the average less well-fitted, because of the 
overwhelming majority of bad over good mutations. Random deaths will only 
occur for a fraction of the population that has more bad mutations than the 
average, and we claim that this strategy brings the model closer to reality. 
It is in fact known that density-dependent components of the demographic 
parameters respond to and affect usually only the numbers of individuals in a 
restricted sub-group of the population, called the \emph{critical age-group} 
\cite{cworth}.

A word of caution is in order here. When simulating the evolution of VB 
populations, one must take special care with the initial transient, which can 
generate populations larger than $Popmax$.

\section{The Penna model}

We will briefly describe in this section the main features of the Penna model 
used in our simulations. For a detailed description of the model, together 
with a complete set of references for work already published on it, we direct 
the reader to Ref. \cite{teubner}.

The genome of each organism is represented by two computer words. In each 
word, a bit set to one at a locus corresponds to a deleterious mutation - a 
``perfect'' strand would be composed solely of zeros. The effect of this 
mutation may be felt by the individual at all ages equal to or above the 
numerical order of that locus in the word. As an example, a bit set to one at 
the second position of one of the bit-strings means that a harmful effect may 
become present in the life history of the organism to which it corresponds 
after it has lived for two time periods. The diploid character of the genome is
related to the effectiveness of the mutations. A mutation in a position of
one of the strands is felt as harmful either because of homozygose or
because of dominance. For the former, a mutation must be present in both
strings at the same position to be effective. The concept of dominance on
the other hand relates to loci in the genome in which a mutation in just one 
strand is enough to make it affect the organism's life. The life span of an 
individual is controlled by the amount of effective mutations active at any 
instant in time. This number must be smaller than a specified threshold to 
keep the individual alive; it dies as soon as this limit is reached. 

Reproduction is modeled by the introduction of new genomes in the
population. Each female becomes reproductive after having reached a
minimum age, after which it generates a fixed number of offspring at the
completion of each period of life. The meiotic cycle is represented by
the generation of a single-stranded cell out of the diploid genome. To do
so, each string of the parent genome is cut at a randomly selected
position, the same for both strings, and the left part of one is combined
with the right part of the other, thus generating two new combinations of
the original genes. The selection of one of these complete the formation
of the haploid gamete coming from the mother. For mating, a male 
is randomly selected in the population and undergoes the same meiotic cycle, 
generating a second haploid gamete out of his genome. The two gametes, one 
from each parent, are now combined to form the genome of the offspring. Each 
of its strands was formed out of a different set of genes. The next stage of 
the reproduction process is the introduction of $M$ independent mutations in 
the newly generated genetic strands. In this kind of model it is normal to 
consider only the possibility of harmful mutations, because of their 
overwhelming majority in nature. The gender of the newborn is
then randomly selected, with equal probability for each sex. 

The passage of time is represented by the reading of a new locus in
the genome of each individual in the population, and the increase of
its age by one. After having accounted for the selection pressure of a
limiting number of effective harmful mutations and the random action of the 
Verhulst dagger, females that have reached the minimum age for reproduction 
generate a number of offspring. The simulation runs for a pre-specified 
number of time steps, at the end of which averages are taken over the 
population. Typically, measures are taken for the age structure of the 
population - number of individuals and probability of survival and death by 
genetic causes for each age group - as well as for the genetic composition 
distribution.

\section{Simulation results}

\begin{figure}[htb]
\centerline{\psfig{file=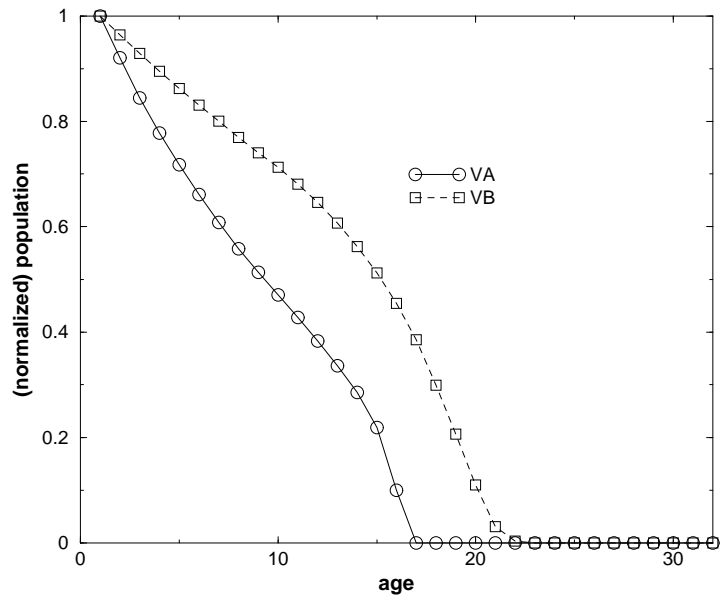, width=8cm, angle=270}}
\caption{Age distribution generated by random deaths over the entire 
population (VA) and only over the newborn (VB). The plot shows the population 
with a certain age, normalized to the population with age $1$ 
($N(age)/N(1)$). Unless otherwise stated, this and the following plots come 
from simulations of iteroparous populations where the genomes were 32-bits 
long, the minimum age at reproduction was $10$, and the threshold for 
deleterious mutations $1$.}
\label{agedis}
\end{figure}

\begin{figure}[htb]
\centerline{\psfig{file=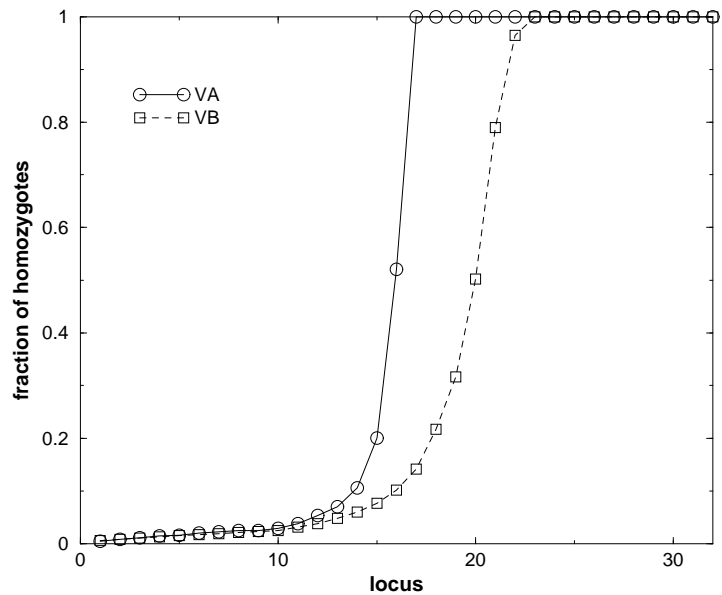, width=8cm, angle=270}}
\caption{The fraction of homozygotes (both bit-strings with a bit set at the 
location) is plotted for each locus in the genome. The simulations were run 
with no dominance, i.e., for a genome to express a defective phenotype, the 
individual had to be an homozygote at a locus.}
\label{fixpat}
\end{figure}

\begin{figure}[htb]
\centerline{\psfig{file=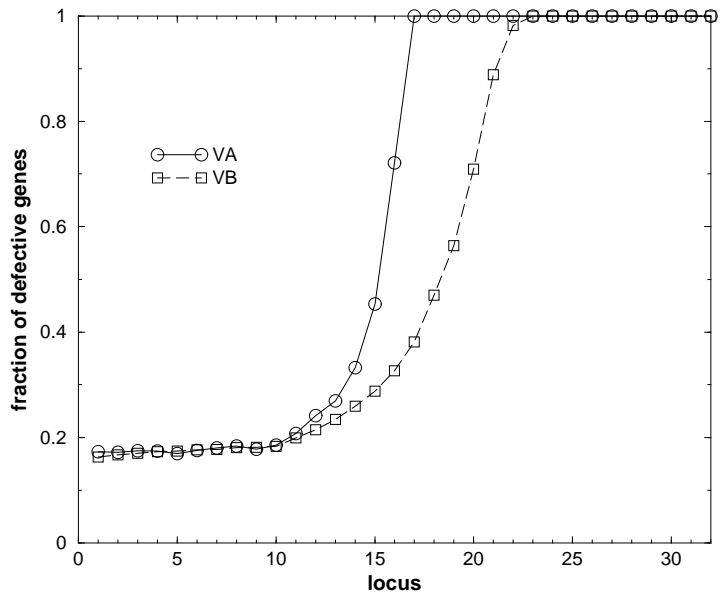, width=8cm, angle=270}}
\caption{The fraction of defective genes (alleles set to one at a location) 
is plotted for each locus in the genome.}
\label{defgen}
\end{figure}

Our claims are supported by the results of simulations performed with the 
bit-string Penna model in which we compare the outcome produced by each of 
the strategies outlined in the last section. First we show that the genetic 
patterns produced by the alternate strategies are not the same. Figure 
\ref{agedis} shows the age distribution generated by both strategies. A 
number of striking differences, apart from the overall concavity of the 
curve, should be noted:
\begin{itemize}
\item The maximum life span is considerably larger for the VB population.
\item The average age of an individual is larger for the VB population.
\item The fraction of the population with reproductive life ($age>10$) is 
also larger for the VB population. As a consequence, the number of offspring 
generated at each time step and the population at equilibrium for the same 
value of the parameter $Popmax$ are also larger.
\end{itemize}
A second comparison is shown in Figure \ref{fixpat}. The pattern of fixation 
of alleles in the genome configuration is shifted upwards in the VB 
population, shrinking the size of the irrelevant (non-selective) part of the 
genome and corresponding to the larger life span shown in the previous plot. 
Figure \ref{defgen} shows yet another feature that is sensitive to the choice 
of random deaths strategy. Here the fraction of defective genes in the 
population is computed for each locus. These last two plots show that the 
genetic configurations of both populations look pretty much the same before 
the age of reproduction. This is not surprising since, for a threshold of 
deleterious mutations of $1$, for a genome to be able to spread throughout 
the population it has to keep the fraction of defective genes and of 
homozygotes at the same low level before the onset of the reproductive period, 
for both populations. During this pre-reproductive period, the random deaths 
are responsible for the faster decrease of the age distribution of the VA 
population (Fig. \ref{agedis}). On the other hand, the genomes show a clear 
distinction in the reproductive period, with a slower rate of degradation for 
the VB population. The better quality of the genomes of the VB population is 
proven by the smaller fraction of deleterious mutations that they carry at 
each age in this period.

The results of our simulation are in fact somewhat surprising. One would 
naively expect, for 
instance, the fraction of random deaths to be larger in a VA population, since 
in this case the Verhulst dagger is allowed to act during the entire life span 
of each individual. Table \ref{rdfrac} shows that this is not the case. In 
fact, simulations show that the fraction of random deaths in the VB population 
is more than twice that for the VA population. The probability of random 
death over the whole life span is also larger for the VB population, in spite 
of only acting on the newborn; for this population, this probability is also 
the fraction of random deaths for the newborn.

A ``coup de grace'' in the set of comparisons we are reporting is the outcome 
of a simulated coevolution of the two populations. We call the reader's 
attention again to numerical problems that may easily be - and have already 
been, at least by one of us! \cite{redq} - overlooked. For any of the 
strategies discussed here, the role of the parameter $Popmax$ has to be 
conveniently downsized. It is very often considered directly in the 
biological sense as a real measure of the capacity of the environment. But we 
argue here that this cannot be so. In fact, in simulations of the case of the 
genes chronologically switched on, with the Verhulst factor killing at each 
time step, the size of a population living in a given environment depends on 
the length of the genome, which has no biological justification. There seems 
to be no biological justification either for a VB population to be of a much 
higher size than a VA population, if both are simulated with the same value 
for $Popmax$, or for an asexual variety to grow larger than a sexual one of 
the same species. In a simulation of the coevolution of VA and VB populations 
without any correction of the $Popmax$ parameter, the VA population gets 
extinct in the first tenths of time steps of coevolution. 
We claim that this is merely an artifact of the model, and that the effective 
carrying capacity of an environment has to be related to the value of the 
population at equilibrium, and cannot be different for different strategies 
of implementation of the concept of random deaths. If one tries to carry a 
realistic coevolution simulation, the parameter $Popmax$ has to be adequately 
manipulated in order to make sure that the size of each population is 
approximately the same at equilibrium, when evolving separately. 

Another important point concerns the stage, during one time step of the 
evolution, at which one has to probe the size of the population for that 
purpose. To understand this point, one has to look back at how these 
simulations are performed. There are some alternatives, but roughly one 
proceeds at each time step sequentially through the population of males and 
females determining the ones that are killed by genetic causes or by random 
deaths, and then mate them to generate the newborn. 
Since we proceed in a sequential manner, at each time step the population 
oscillates between a minimum value, which is its value after all deaths for 
that particular time step have been considered, and a maximum one at the end 
of the time step, after all births. These cycles of compression (deaths) and 
expansion (births) are not real, but merely artifacts of a sequential 
processing.
It is usual to compute the total population at the beginning of a time step. 
This means that it includes all the individuals of $age > 0$ that remained 
alive after the completion of the last time step \emph{plus} the newborn. 
However, as above explained, this is a \emph{peak} value for a fluctuating 
population, and can never actually be seen, for the dying and breeding 
processes happen in parallel in reality. 
The actual population for the purposes of comparison with the real carrying 
capacity of an environment cannot be taken at this peak value, but rather at 
some smaller one. Since the pressure of the newborn over the capacity of the
environment can be neglected in the presence of that of the individuals with 
$age > 0$, we chose to pick the minimum value of the population at each time 
step as the \emph{real} one. 

\begin{figure}[htb]
\centerline{\psfig{file=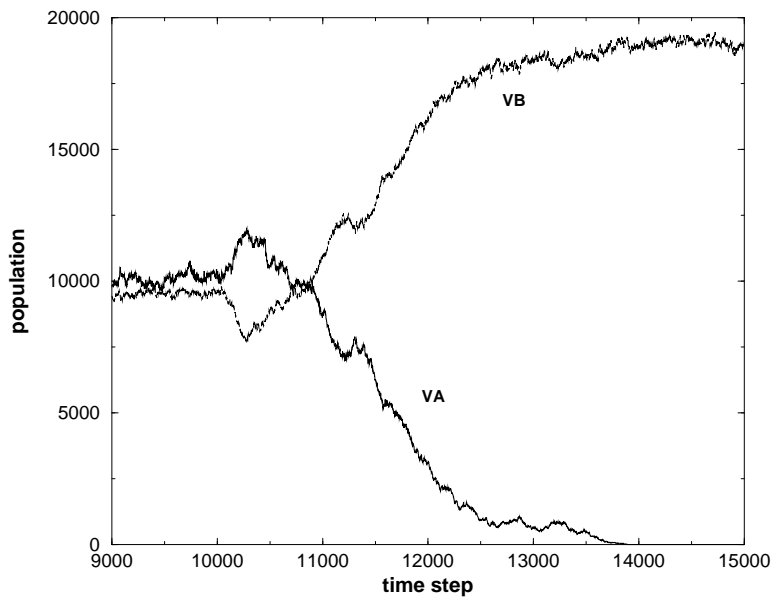, width=8cm, angle=270}}
\caption{The population for each variety is plotted against time. Coevolution 
starts at time step $10,000$, after a sufficiently long transient. For the 
parameters we are using, in particular $Popmax=200,000$, the compression 
factor to be applied for the VB population to ensure equal populations of 
$\approx 10,000$ in equilibrium is $10.5$, leading to an effective value of 
$Popmax \approx 19,000$ for this population; in fact, this factor is a little 
bit too large, as can be seen in the plot through a slightly larger VA 
population at the beginning. In less than $5,000$ steps the VA population 
dies out.}
\label{coev}
\end{figure}

Once these considerations are taken into account, the parameter $Popmax$ has 
to be compressed by some factor for the VB population to ensure that its size 
at equilibrium matches that of the VA population, all the other parameters 
being equal. Figure \ref{coev} shows the result of simulating the 
coevolution of the two populations. After having evolved independently for 
some time, to wash out any transient behaviour, the populations are brought 
into contact, sharing the same environment. The VA population dies out after 
less than $5,000$ time steps. This outcome is a convincing support for the 
claims previously made. Nature would, if necessary, rather sacrifice those 
upon whom less investment has been made, as it often does in animal 
populations.

\begin{figure}[htb]
\centerline{\psfig{file=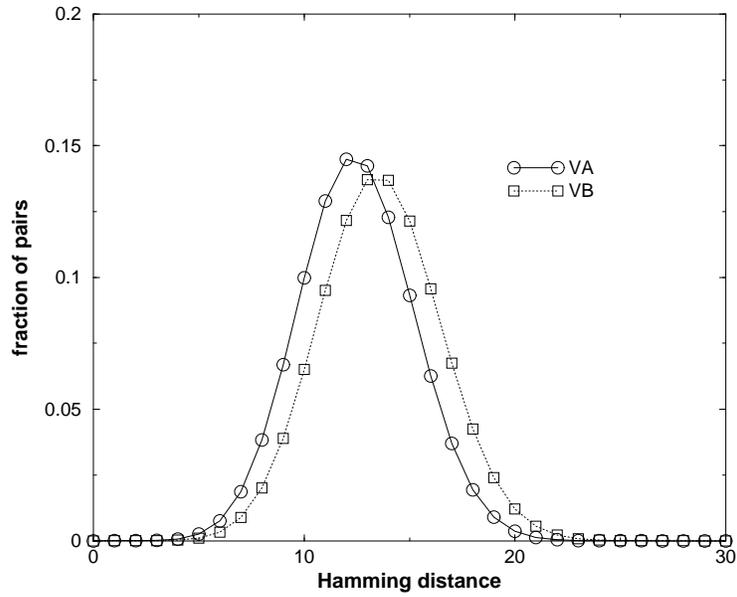, width=8cm, angle=270}}
\caption{Histogram of Hamming distances for both populations. The plot shows 
the fraction of all pairings that share each value for the distance.}
\label{hamm}
\end{figure}

We address finally the question of diversity. Would a VB strategy decrease 
the genetic diversity provided by sexual reproduction? One might think so, 
since random deaths are concentrated on a fraction of the population with 
a larger number of mutations. Figure \ref{hamm} supports a negative answer. 
Diversity is measured through a histogram of Hamming distances across the 
population, defined as the number of different alleles (bits) in every pair 
of genomes \cite{why}. The VB strategy even gives diversity a small 
enhancement. 

\section{Conclusions}

In Nature, at least in the case of higher diploid organisms, whose 
populations are simulated by the Penna model, random deaths play no 
significant role, and computational models 
that want to capture the essence of evolution must take this into account. 
We have shown that the choice of strategy in implementing this concept, 
which is unfortunately necessary to prevent unlimited growth of the 
population in those models, has an unexpected impact on its genetic profile. 
We claim that the choice of exposing only the newborn to random deaths is at 
present the most realistic one, and limits the aforementioned impact to a 
minimum. As an illustration, we presented an evolutionary argument based on 
the simulated competition of two non-crossing varieties of the same species 
where the outcome shows that Nature would most probably choose the same 
strategy as we did. It is our impression that some of the results obtained 
with the use of the models that rely on random deaths for their stabilization 
should be revised in light of the present discussion.

\section*{Acknowledgments}
D. Stauffer was responsible for arranging our meeting in the virtual space; 
we want to thank him for that and for his encouragement and intellectual 
support.
J.S.S.M. was supported by DOE grant DE-FG03-95ER14499 and C.S. by UW grant 
2027/W/Imi/2000.

\newpage

\begin{table}
\caption{Births and random deaths for both strategies. In the last row, we 
show the probability of dying by the Verhulst dagger over the entire life 
span.}
\begin{tabular}[t]{l|c|c}
& VA & VB \\
\hline
births/population & 0.111 & 0.167 \\
random deaths/population & 0.043 & 0.097 \\
random deaths/births & 0.387 & 0.580 \\
probability of random death & 0.443 & 0.580 \\
\hline
\label{rdfrac}
\end{tabular}
\end{table} 

\end{document}